\documentclass{article}
\usepackage{spconf,amsmath,amssymb,graphicx,hyperref}
\usepackage{multirow}
\usepackage{booktabs}
\usepackage{bm}
\usepackage[table]{xcolor}
\usepackage{enumitem}
\definecolor{lightyellow}{HTML}{FEF3D4} 
\title{MeanFlow-Accelerated Multimodal Video-to-Audio Synthesis via One-Step Generation}

\name{Xiaoran Yang$^{1\dagger}$, Jianxuan Yang$^2$, Xinyue Guo$^2$, Haoyu Wang$^{3\dagger}$, Ningning Pan$^3$, Gongping Huang$^{1*}$ \thanks{$^{\dagger}$ Work done during an internship at Xiaomi.} \thanks{* Corresponding Author.} }

\address{ \large
\hskip -7pt \begin{tabular}{c}
$^1$School of Electronic Information, Wuhan University, Wuhan, China\\
$^2$MiLM Plus, Xiaomi Inc., Wuhan, China\\
$^3$School of Computing and Artificial Intelligence,\\
Southwestern University of Finance and Economics, Chengdu, China
\end{tabular}
}

\begin{document}
\ninept

\maketitle

\begin{abstract}
A key challenge in synthesizing audios from silent videos is the inherent trade-off between synthesis quality and inference efficiency in existing methods. For instance, flow matching based models rely on modeling instantaneous velocity, inherently require an iterative sampling process, leading to slow inference speeds. 
To address this efficiency bottleneck, we introduce a MeanFlow-accelerated model that characterizes flow fields using average velocity, enabling one-step generation and thereby significantly accelerating multimodal video-to-audio (VTA) synthesis while preserving audio quality, semantic alignment, and temporal synchronization. 
Furthermore, a scalar rescaling mechanism is employed to balance conditional and unconditional predictions when classifier-free guidance (CFG) is applied, effectively mitigating CFG-induced distortions in one step generation.
Since the audio synthesis network is jointly trained with multimodal conditions, we further evaluate it on text-to-audio (TTA) synthesis task. 
Experimental results demonstrate that incorporating MeanFlow into the network significantly improves inference speed without compromising perceptual quality on both VTA and TTA synthesis tasks. Demos are provided in \url{https://vta888.github.io/MF-MJT/}  
\end{abstract}


\section{Introduction}
\label{sec:intro}

Video-to-audio (VTA) synthesis focus on generating semantically aligned and temporally synchronized audios given silent video inputs, with applications such as video dubbing. Existing approaches can be broadly divided into three paradigms. The first trains models solely on paired audio-visual data from scratch~\cite{wang2024frieren, Cheng2025LoVA}. However, paired datasets are typically scarce and often of limited qualit. 
The second leverages pretrained text-to-audio (TTA) synthesis models~\cite{liu2023audioldm, girdhar2023imagebind} by adding auxiliary modules that adapt to visual inputs~\cite{xing24seeing, zhang2024foleycrafter}. Nevertheless, it remains unclear whether the additional modules can fully bridge the inherent domain gap between text and video, and the reliance on pretrained models constrains flexibility. 
To overcome these limitations, a third paradigm, namely multimodal joint training, has recently emerged~\cite{You2025TAV2A, cheng2025taming, Jun2025kling}. This framework learns from scratch by jointly considering video, audio, and text conditions in an end-to-end manner, which enables scalable data utilization and establishes a unified semantic space that facilitates cross-modal understanding. In this work, we adopt multimodal joint training framework as the backbone of our VTA synthesis network.

\begin{figure}[t!]
\centering
\includegraphics[width=\linewidth]{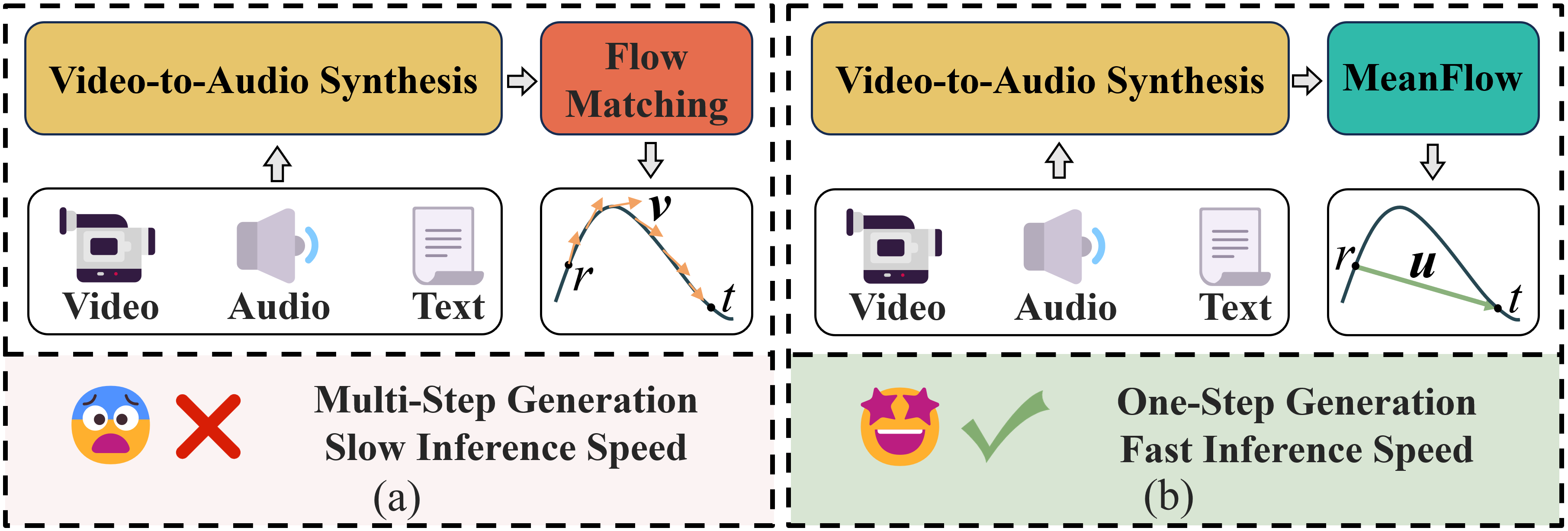}
 \vspace{-0.7cm}
    \caption{Comparison of VTA synthesis approaches: (a) previous FM based method modeling instantaneous velocity, and (b) proposed MeanFlow-accelerated method modeling average velocity.}
\label{fig:fig1}
\end{figure}

An effective VTA synthesis network is expected to excel in four critical points: semantic alignment, temporal synchronization, audio quality, and inference efficiency, among which inference efficiency remains the bottleneck.
Such diffusion based~\cite{Ho2020ddpm} or flow matching (FM) based~\cite{liu2022flow} generative models often require iterative sampling steps to approximate the data distribution. For instance, FM based methods model the instantaneous velocity and thereby require multiple sampling steps, resulting in slow inference, as illustrated in Fig.~\ref{fig:fig1} (a). 
To improve efficiency, Frieren~\cite{wang2024frieren} adopts rectified FM~\cite{liu2022flow} and one-step distillation~\cite{liu2023instaflow}, but still suffers from costly multi-stage training, multiple steps for quality, and reliance on pretrained modules.
\cite{Geng2025meanflow} proposes mean flows that directly characterizes the flow field with average velocity, enabling native one-step generation without auxiliary distillation, which is adopted in our work, as illustrated in Fig~\ref{fig:fig1} (b).
Moreover, to further enhance flexibility, classifier-free guidance (CFG) is commonly employed by combining conditional and unconditional predictions~\cite{ho2022classifier}. However, it may cause overshooting due to the lack of iterative refinement in the one-step generation. 
Recent research CFG-Zero~\cite{fan2025cfgzerostar} shows that introducing an optimized scalar can effectively mitigates this issue by balancing the contributions of conditional and unconditional predictions. 

Thus, to improve both efficiency and flexibility for VTA synthesis, we propose a MeanFlow-accelerated multimodal joint training network (MF-MJT), incorporating a CFG-scaled
mechanism.
The main contributions are: (i) we introduce a MeanFlow-accelerated framework for multimodal video-to-audio synthesis that enables native one-step generation and significantly improves efficiency; (ii) we propose a CFG-scaled mechanism to stabilize CFG within MeanFlow, effectively enhancing perceptual quality; and (iii) we conduct extensive experiments on VTA and TTA synthesis tasks, demonstrating substantial acceleration, ranging from $2\times$ to $500\times$, without compromising perceptual quality.

\section{method}
\label{sec:met}

The proposed MF-MJT builds upon a joint training backbone, known as MMAudio~\cite{cheng2025taming}, that integrates video, audio, and text modalities within a unified framework. On top of this backbone, MeanFlow formulation is introduced which directly models the average velocity to enable native one-step generation. Furthermore, to stabilize CFG under one-step generation, a scalar rescaling mechanism is employed to balance conditional and unconditional predictions. This section first introduces the backbone architecture of MF-MJT, followed by a detailed discussion of the MeanFlow-accelerated and improved CFG strategies for efficient inference.

\subsection{Multimodal Joint Training Backbone}
\label{ssec:MJT}

\begin{figure}[t!]
\centering
\includegraphics[width=\linewidth]{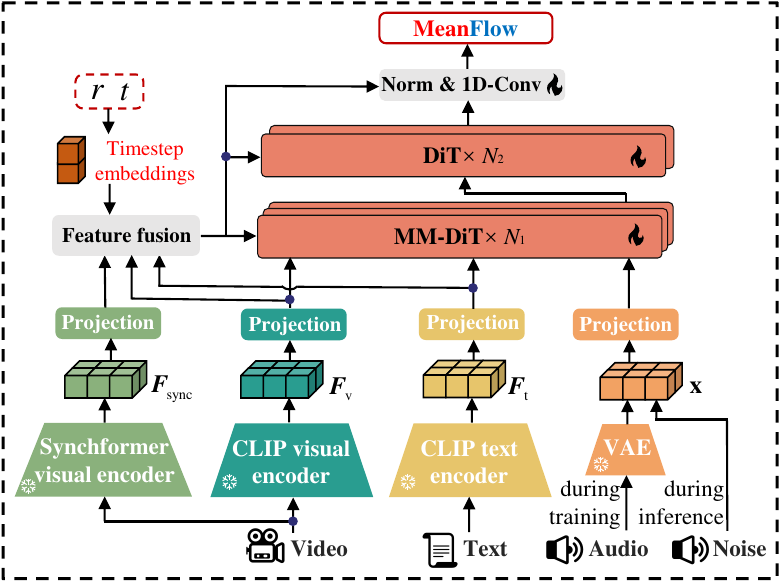}
 \vspace{-0.7cm}
    \caption{The model architecture of the proposed Meanflow-accelerated VTA synthesis network, MF-MJT.}
\label{fig:fig2}
\end{figure}

The model architecture is illustrated in Fig.~\ref{fig:fig2}, which follows the mainstream Multimodal Diffusion Transformer (MM-DiT)~\cite{esser2024scal} combined with the Diffusion Transformer (DiT)~\cite{Peebles2023Dit} design. The MM-DiT blocks take three types of inputs: video, text, and audio. Specifically, the visual features $\bm F_\mathrm v$ and text features $\bm F_\mathrm t$ are extracted using CLIP encoders~\cite{Radford2021LearningTV}, while the audio latents $\mathbf{x}$ are obtained from a pretrained Variational Autoencoder (VAE)~\cite{Kim2025vae} during training and replaced with random noise during inference. These embeddings are then projected into a unified latent space and fused through cross-modal attention layers, enabling effective alignment across modalities. In addition, a Synchformer visual encoder~\cite{Iashin2024sync} is employed to enhance audio-visual synchrony. Its output features $\bm F_\mathrm {sync}$, together with $\bm F_\mathrm v$ and $\bm F_\mathrm t$, are projected and pooled before being fused with the timestep embeddings to provide frame-aligned conditional control for generation.

After the joint-attention outputs of $N_1$ MM-DiT block are partitioned back into the three modalities, the audio branch is further processed by $N_2$ audio-only DiT blocks, where the cross-modal attention is replaced by self-attention to refine audio-specific representations. Finally, the audio latents are processed through adaptive layer normalization followed by a 1D convolution layer, yielding the estimated average velocity field $\bm u_\theta$.

\subsection{MeanFlow-accelerated One-step Generation}
\newcommand{\utgt}{\bm u_\text{tgt}}

\label{ssec:MF}

FM~\cite{liu2022flow} formulates generative modeling as learning an ordinary differential equation (ODE)~\cite{Yang2021scoresde} that transports a prior sample $\bm {\epsilon} \sim p_{\text{prior}}(\bm {\epsilon})$ to a target $\bm x \sim p_\text{data}(\bm x)$ by estimating the instantaneous velocity field $\bm v(\bm z_t, t)$ along the flow path $\bm z_t = (1-t)\bm x + t\bm {\epsilon}$ with time $t$. The ground-truth velocity is $\bm v(\bm z_t, t) = \bm z'_t =\bm {\epsilon} - \bm x$, and a neural network $\bm v_\theta$ is trained to minimize:
\begin{align}
\mathcal{L}_\text{FM}(\theta) = \mathbb{E}_{t,\bm x,\bm {\epsilon}} \| \bm v_\theta(\bm z_t, t) - \bm v(\bm z_t, t)) \|^2.
\label{eq:loss_fm}
\end{align}
The sampling process then solves the ODE:
\begin{align}
\frac{d\bm z_t}{dt} = \bm v_\theta(\bm z_t, t),
\end{align}
which requires iterative integration from $\bm z_1$ to $\bm z_0$, resulting in high computational cost and slow inference.

To overcome this limitation, we introduce the MeanFlow formulation~\cite{Geng2025meanflow} into MF-MJT. It replaces the local instantaneous velocity with a global average velocity, which is defined as:
\begin{align}
\bm u(\bm z_t, r, t) =  \frac{1}{t-r} \int_{r}^{t}\bm v(\bm z_\tau, \tau)d\tau,
\label{eq:average_u}
\end{align}
where 
$(r,t)$ are two time steps with $t \geq r$. This formulation provides an alternative ground-truth field for learning.
By differentiating Eq.~\eqref{eq:average_u} with respect to $t$, we obtain the MeanFlow identity:
\begin{align}
\bm u(\bm z_t, r, t) =\bm v(\bm z_t, t)- (t - r)\frac{d}{dt}\bm u(\bm z_t, r, t).
\label{eq:identity}
\end{align}
A neural network $\bm u_\theta$ is then trained to satisfy this identity by minimizing:
\begin{align}
\mathcal{L}_\text{MF}(\theta) & = \mathbb{E}_{r,t,\bm x,\bm {\epsilon}} \big\| \bm u_\theta(\bm z_t, r, t) - \text{sg}(\utgt) \big\|_2^2, 
\label{eq:meanflow_loss} 
\end{align}
where $\utgt = \bm v_t - (t - r)
\big( 
\bm v_t\partial_{\bm z}{\bm u_{\theta}}+
\partial_{t}{\bm u_{\theta}}\big)$ and the operation $\text{sg}(\cdot)$ denotes stop-gradient. Notably, this objective reduces to the standard FM loss when $r = t$.

During inference, MeanFlow enables direct mapping
by replacing time integral with the average velocity: 
\begin{align}
 \bm z_r = \bm z_t - (t-r)\bm u(\bm z_t,r,t).
\label{eq:sample_rt} 
\end{align}
This formulation naturally supports multi-step sampling in a straightforward manner, as demonstrated in Section~\ref{sec:Res}. For the one-step case, which is the primary focus of this work, Eq.~\eqref{eq:sample_rt} simplifies to $\bm z_{0} = \bm z_{1} - \bm u(\bm z_{1}, 0, 1)$,
where $\bm z_{1}=\bm {\epsilon} \sim p_\text{prior}(\bm {\epsilon})$. 
It achieves efficient one-step generation without iterative refinement, but applying strong classifier-free guidance under this setting introduces new challenges, which we address in the next subsection.

\subsection{CFG with Scalar Rescaling}
\label{ssec:cfg}

CFG is commonly employed to enhance conditional generation quality by combining conditional and unconditional predictions~\cite{ho2022classifier}. During training, the conditions $\mathbf{c}$, including $\bm F_\mathrm v$, $\bm F_\mathrm t$ and $\bm F_\mathrm {sync}$, are randomly dropped with a probability of 0.1. At inference, the vector field is modified as:
\begin{align}
\bm u^\text{cfg}_\theta = \omega \cdot \bm u_\theta(\bm z_t, r, t \mid \mathbf{c}) + (1 - \omega) \cdot \bm u_\theta(\bm z_t, r, t \mid \bm \varnothing),
\label{eq:cfg}
\end{align}
where $\omega$ denotes the guidance strength. However, in a one-step generation setting, an inappropriate $\omega$ may cause overshooting and noticeable artifacts due to the lack of iterative refinement.

To alleviate this issue, we incorporate a scalar rescaling strategy, which prevents CFG from steering samples along incorrect trajectories and corrects guidance inaccuracies:
\begin{align}
\bm u^\text{cfg-scaled}_\theta = \omega \cdot \bm u_\theta(\bm z_t, r, t \mid \mathbf{c}) + (1 - \omega) \cdot s  \cdot \bm u_\theta(\bm z_t, r, t \mid \bm \varnothing),
\label{eq:scaled_cfg}
\end{align}
where
\begin{align}
s = \frac{\bm u_\theta(\bm z_t, r, t \mid \mathbf{c})^\top \bm u_\theta(\bm z_t, r, t \mid \bm \varnothing)}{\left\lVert \bm u_\theta(\bm z_t, r, t \mid \bm \varnothing) \right\rVert^2}
\label{eq:scaled_s}
\end{align}
is an adaptive scalar of the conditional velocity projected onto the direction of the unconditional velocity, reflecting their alignment. By dynamically adjusting the unconditional component, the guided update direction remains semantically consistent with the conditional prediction, thereby avoiding deviations from the correct trajectory.
This simple yet effective adjustment stabilizes guidance in one-step generation and significantly improves perceptual quality.

\section{Experimental Setup}
\label{sec:exp}

\renewcommand{\arraystretch}{1}
\begin{table*}[t!]
\centering
\caption{The performance of Video-to-Audio synthesis methods on the VGGSound test set. All samples are generated on an H800 GPU. The best results for each metric in both single-step and multi-step scenarios are indicated in bold respectively. }
\vspace{4pt}
\setlength{\tabcolsep}{7pt} 
\begin{tabular}{lccccccccc>{\columncolor{lightyellow}}c}
\hline
\multirow{2}{*}{\textbf{Method}} & \multirow{2}{*}{STEP} & \multirow{2}{*}{Params} & \multicolumn{3}{c}{Distribution matching} & \multicolumn{1}{c}{Audio quality} & \multicolumn{1}{c}{Semantic align} & \multicolumn{1}{c}{Temporal Sync} & Efficiency  \\
\cmidrule(lr){4-6}
\cmidrule(lr){7-7}
\cmidrule(lr){8-8}
\cmidrule(lr){9-9}
\cmidrule(lr){10-10}
& & & FAD~\(\downarrow\) & FD~\(\downarrow\) & KL~\(\downarrow\) & IS~\(\uparrow\) & IB~\(\uparrow\) & DeSync~\(\downarrow\) & RTF~\(\downarrow\)\\
\hline
Frieren  & \multirow{2}{*}{1}  &  159M & 1.87 & 16.64 & 2.56 & 9.14 &  \textbf{21.92} & \textbf{0.85} &  \cellcolor{lightyellow}0.015 \\
MF-MJT (ours) & &  157M & \textbf{1.46} &\textbf{11.14} & \textbf{1.87} & \textbf{9.39} &  21.78 &  0.86 & \cellcolor{lightyellow}\textbf{0.007} \\
\hline
Seeing\&Hearing & \multirow{5}{*}{25} & 415M & 4.69 & 29.00 & 2.92 & 6.16 & \textbf{33.85} & 1.19 & \cellcolor{lightyellow}3.772  \\
FoleyCrafter & & 1.22B & 2.60 & 16.25 & 2.31 & 16.03 & 25.75 & 1.22 & \cellcolor{lightyellow}0.463 \\ 
Frieren & & 159M  & 1.35 & 11.54 & 2.75 & 13.06 & 22.90 & 0.85 & \cellcolor{lightyellow}0.335 \\
MMAudio & & 157M  & \textbf{0.81} & \textbf{4.79} & 1.63 & 13.43 & 28.69 & \textbf{0.51} & \cellcolor{lightyellow}\textbf{0.098} \\
MF-MJT (ours) & & 157M   & 1.13 & 5.87 & \textbf{1.59} & \textbf{16.55} & 28.22 & 0.57 & \cellcolor{lightyellow} 0.101 \\
\hline
\end{tabular}
\label{tab:vta}

\end{table*}
\renewcommand{\arraystretch}{1}

\subsection{Multimodal Dataset}

The proposed MF-MJT is trained on multimodal datasets comprising both audio-video-text and audio-text pairs. Specifically, VGGSound (approximately 500 hours) ~\cite{Chen2020vgg} and Kling-Audio-Eval (around 58 hours)~\cite{Jun2025kling} provide audio-video-text triplets. To expand the training data coverage, we also include audio-text datasets such as AudioCaps (approximately 128 hours)~\cite{kim-etal-2019-audiocaps} and WavCaps (about 7,600 hours)~\cite{mei2023wavcaps}, where the corresponding visual and synchronization features are represented by empty tokens \(\bm \varnothing_{\mathrm{v}}\) and \(\bm \varnothing_{\mathrm{sync}}\). 
For evaluation, the VTA synthesis performance is assessed on the VGGSound test sets including 15,216 samples, while the TTA synthesis performance is evaluated on the AudioCaps test set comprising 4,227 samples.

\subsection{Implementation Details}
\textbf{Model Variants.} The numbers of MM-DiT and DiT blocks are set to $N_1 = 4$ and $N_2 = 8$, respectively. Audios are sampled at 31.25 frames per second (fps) and embedded as 20-dimensional audio latents $\mathbf{x}$. The visual features $\bm F_\mathrm v$ (one token per frame at 8 fps) and text features $\bm F_\mathrm t$ (77 tokens) are encoded into 1024-dimensional representations, while the synchronization features $\bm F_\mathrm {sync}$ are extracted at 24 fps as 768-dimensional features. 

\noindent \textbf{MeanFlow Formulation.} The two time steps $(r,~t)$ are sampled from a logit-normal distribution, where each sample is drawn from a normal distribution $\mathcal{N}(\mu,\sigma)$ and subsequently mapped to the range $(0, 1)$ via the logistic function. In this work, both $r$ and $t$ adopt $\mu = -2.0$ and $\sigma = 2.0$, with the constraint $r \le t$. The ratio of $r \neq t$ is set to $10\%$, as later experiments demonstrate that a lower ratio leads to better performance. Positional embeddings are used to encode the time variables, which are fused and provided as conditioning inputs to the generation network. Notably, the network does not condition directly on $(r,~t)$, but rather on $(t, ~\Delta t)$, where $\Delta t = t - r$.

\noindent \textbf{Training and Inference.} We trained our model on eight NVIDIA H800 GPUs with a batch size of $64$ per device for $400,000$ steps on the aforementioned multimodal dataset. We employ the AdamW optimizer~\cite{loshchilov2019decoupled} with $\beta_1=0.9$ and $\beta_2=0.95$, a maximum learning rate of $2 \times 10^{-4}$, and a weight decay of $1 \times 10^{-6}$. The learning rate is linearly warmed up during the first $1{,}000$ steps, then decayed to $2 \times 10^{-5}$ at step $250{,}000$ and further to $2 \times 10^{-6}$ at step $350{,}000$. For inference, we perform one-step generation by fixing $(r,~t)=(0,~1)$ and apply classifier-free guidance with a strength of $1.5$. Notably, absolute positional encodings are omitted, enabling the model to synthesis variable-length audio (e.g., $8$~s for VGGSound and $10$~s for AudioCaps).

\subsection{Evaluation Metrics}
We evaluate the generation performance across multiple dimensions: distribution matching, audio quality, semantic alignment and efficiency for both VTA and TTA synthesis task, with an additional focus on temporal synchronization for VTA.
Distribution matching is assessed using Fréchet audio distance (FAD)~\cite{Gemmeke2017Audio}, Fréchet distance (FD)~\cite{Kong2020panns}, and Kullback–Leibler divergence (KL)~\cite{Kong2020panns}, where FAD is computed with VGGish embeddings, and FD and KL are based on PANNs representations. Audio quality is measured using the inception score (IS)~\cite{koutini22passt} with a PaSST classifier, while semantic alignment is evaluated by the ImageBind (IB) score~\cite{girdhar2023imagebind} for VTA and the CLAP score~\cite{Elizalde2023CLAP} for TTA. The temporal synchronization is quantified by the DeSync score~\cite{Iashin2024sync} for VTA. To assess efficiency, the real-time factor (RTF), representing the ratio of inference time to generated audio duration on a single NVIDIA H800 GPU, is calculated and averaged across the entire test set.

\subsection{Baselines}
For VTA synthesis, we compare the proposed MF-MJT with four representative baselines: Foleycrafter~\cite{zhang2024foleycrafter}, Seeing and Hearing~\cite{xing24seeing}, MMAudio~\cite{cheng2025taming}, and Frieren~\cite{wang2024frieren}. The first three methods rely on multi-step generation, whereas Frieren supports both single-step and multi-step generation. For TTA synthesis, we evaluate MF-MJT against AudioLCM~\cite{Liu2024AudioLCM} and MMAudio. Specifically, AudioLCM is also designed for efficient inference. 
Notebly, we adopt their recommended default CFG strength for all baseline methods, and the number of sampling steps for all multi-step approaches is fixed at $25$ in our experiments to ensure a fair comparison of RTF. For MF-MJT, the CFG strength is set to $1.5$ for single-step generation and $4.5$ for multi-step generation to achieve optimal performance.

\section{Results}
\label{sec:Res}

\subsection{Comparison with Baselines}

Table~\ref{tab:vta} summarizes the performance of the proposed MF-MJT against representative VTA synthesis baselines on the VGGSound test set. For one-step generation, the MF-MJT outperforms Frieren on the majority of metrics, demonstrating its ability to generate perceptually high-quality audio across diverse evaluation dimensions. For multi-step generation, the MF-MJT still remains highly competitive, achieving the best performance on KL and IS, while other metrics are on par with MMAudio. Notably, the MF-MJT exhibits remarkable inference efficiency: the RTF for one-step generation is only 0.007, significantly lower than all other methods, while multi-step generation remains fast with an RTF of 0.101, indicating that the proposed approach offers flexibility in the number of sampling steps rather than being limited to single-step generation. This efficiency gain highlights the advantage of the MeanFlow-accelerated one-step formulation, enabling rapid, high-fidelity video-to-audio synthesis suitable for practical applications.

\begin{table}[t!]
\centering
\caption{The performance of Text-to-Audio synthesis methods on the AudioCaps test set. All samples are generated on an H800 GPU. The best results for each metric oth single-step and multi-step scenarios are indicated in bold respectively. }
\vspace{4pt}
\setlength{\tabcolsep}{3.5pt} 
\begin{tabular}{lcccccc}
\hline
Method & STEP & FAD~\(\downarrow\) & FD~\(\downarrow\) &  IS~\(\uparrow\) & CLAP~\(\uparrow\) & RTF~\(\downarrow\) \\
\hline
AudioLCM  & \multirow{2}{*}{1} & 4.24 & 23.16 &  \textbf{7.13} & 0.19 &\cellcolor{lightyellow} 0.016\\
MF-MJT (ours) & &  \textbf{2.29} &  \textbf{21.32} & 6.50 &  \textbf{0.20} &  \cellcolor{lightyellow}\textbf{0.007}  \\
\hline
AudioLCM & \multirow{3}{*}{25}  & \textbf{1.79} & 16.91 & 9.77 & 0.24 &  \cellcolor{lightyellow} \textbf{0.056} \\
MMAudio & & 2.96 & 14.51 & 11.29 & 0.28 & \cellcolor{lightyellow} 0.085  \\
MF-MJT (ours) & & 2.83 & \textbf{14.06} & \textbf{11.53} & \textbf{0.30} & \cellcolor{lightyellow} 0.086 \\
\hline
\end{tabular}
\label{tab:tta}
\end{table}

Then, the MF-MJT is applied to TTA synthesis task without fine-tuning. Table~\ref{tab:tta} presents its performance compared with other baselines on the AudioCaps test set. In the one-step setting, MF-MJT surpasses AudioLCM on FAD, FD, and CLAP while achieving an extremely low inference cost (RTF = 0.007), demonstrating its efficiency and effectiveness. For multi-step generation, MF-MJT achieves the best performance on FD, IS, and CLAP, outperforming MMAudio. These results highlight that MF-MJT delivers strong performance in both one-step and multi-step settings, offering a favorable balance between quality and efficiency for the TTA task.

\subsection{Ablations}

\begin{figure}[t!]
\centering
\includegraphics[width=\linewidth]{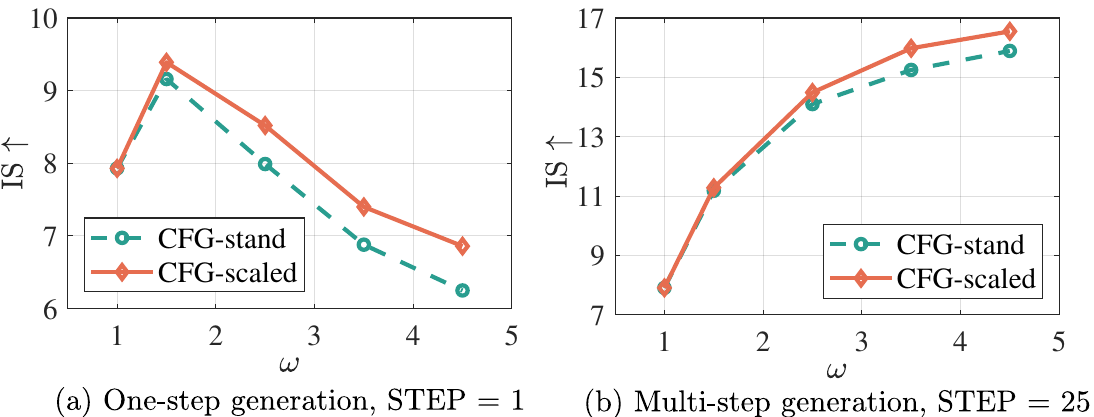}
 \vspace{-0.7cm}
    \caption{Comparison of the standard CFG and the proposed CFG-scaled strategies as a function of CFG strength, $\omega$, on the VGGSound test set.} 
\label{fig:cfg}
\end{figure}

We first investigate whether the proposed scalar rescaling mechanism, CFG-scaled, is effective within the MeanFlow-Accelerated framework on the VTA synthesis task. Figure~\ref{fig:cfg} (a) and (b) present the IS, reflecting the perceptual quality of audio, with and without CFG-scaled under one-step and multi-step settings, respectively. 
The CFG strength $\omega$ is set to $1.0$, $1.5$, $2.5$, $3.5$, and $4.5$. When $\omega=1.0$, CFG is disabled. 
It is observed that MF-MJT 
CFG-scaled consistently outperforms that using standard CFG (CFG-stand) across different CFG strength values $\omega$ (except when $\omega = 1.0$).
Moreover, in the one-step generation setting, the performance of MF-MJT tends to degrade as $\omega$ increases, whereas the opposite trend is observed under multi-step generation. This is reasonable because one-step generation lacks an iterative refinement mechanism—an excessively strong guidance may cause the update direction to deviate significantly from the optimal trajectory. In contrast, multi-step generation can gradually correct such deviations in subsequent steps, even when $\omega$ is relatively large.

\begin{figure}[t!]
\centering
\includegraphics[width=\linewidth]{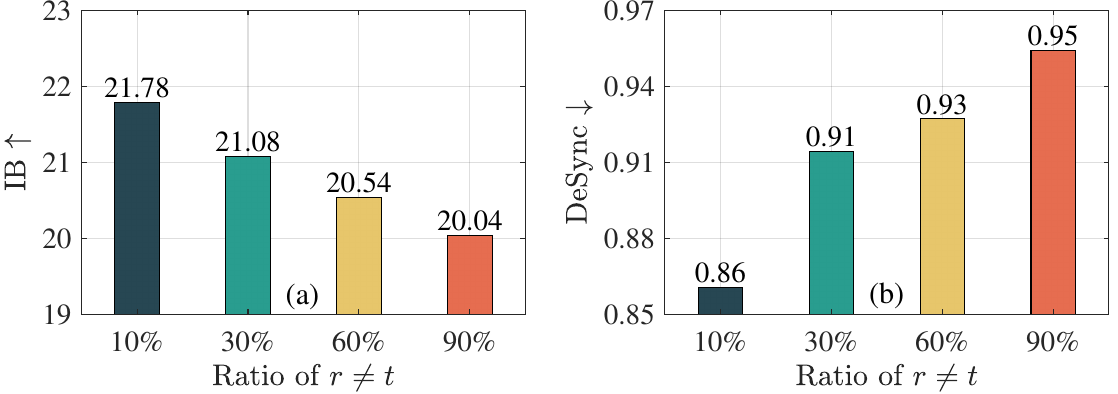}
 \vspace{-0.7cm}
    \caption{Impact of the ratio of $r \neq t$ pairs during training, evaluated on the VGGSound test set under one-step generation. (a) IB score. (b) DeSync score. $\omega = 1.5$.}
\label{fig:rt_ratio}
\end{figure}

We further explore the role of sampling pairs where $r \neq t$ in the MeanFlow formulation during the training stage. To assess the impact, we consider two representative metrics: IB and DeSync scores, evaluating the semantic alignment and temporal synchronization between audio and video, respectively. As illustrated in Fig.~\ref{fig:rt_ratio}, a lower ratio of $r \neq t$ pairs leads to consistent improvements in both two metrics. This phenomenon can be attributed to the role of $r=t$ pairs in providing direct supervision for aligning audio and video representations at the same temporal position, thereby facilitating precise cross-modal alignment. In contrast, an excessive inclusion of $r \neq t$ pairs introduces ambiguous or conflicting contrastive signals that may confuse the network.

\section{Conclusion}
\label{sec:Conc}

We propose MF-MJT, the first MeanFlow-accelerated framework for multimodal video-to-audio synthesis, enabling native one-step generation by modeling average velocity. To enhance the controllability of classifier-free guidance, we introduce a scalar rescaling mechanism (CFG-scaled) that mitigates overshooting without iterative refinement and further improves the perceptual quality of the synthesized audio. Experiments on VGGSound and AudioCaps test sets demonstrate that MF-MJT achieves an RTF of $0.007$ for one-step inference on an NVIDIA H800 GPU under both VTA and TTA synthesis tasks. 
This delivers over $2 \times$ speedup compared to baseline approaches, including other one-step generation methods. Despite this significant acceleration, MF-MJT maintains comparable audio quality, semantic alignment, and temporal synchronization. 
Even in multi-step settings, MF-MJT remains competitive across all evaluation metrics, demonstrating strong generalization for both VTA and TTA synthesis tasks.

\bibliographystyle{ieeetr}
\bibliography{refs}

\end{document}